\documentclass[twocolumn,showpacs,floatfix,superscriptaddress,amsmath,amssymb,prl]{revtex4}
\usepackage{mathrsfs}
\usepackage{txfonts}
\usepackage{amssymb}
\usepackage{graphicx}
\usepackage{hyperref}
\usepackage{ulem}
 \usepackage{overpic}
 \usepackage{psfrag}
 \newcommand{\be}{\begin{equation}}
\newcommand{\ee}{\end{equation}}

\begin{document}
\title{Kosterlitz-Thouless Phase Transition and Ground State Fidelity: a Novel Perspective from Matrix Product States}

\author{Hong-Lei Wang}
\affiliation{Centre for Modern Physics and Department of Physics,
Chongqing University, Chongqing 400044, The People's Republic of
China}
\author{Jian-Hui Zhao}
\affiliation{Centre for Modern Physics and Department of Physics,
Chongqing University, Chongqing 400044, The People's Republic of
China}
\author{Bo Li}
\affiliation{Centre for Modern Physics and Department of Physics,
Chongqing University, Chongqing 400044, The People's Republic of
China}
\author{Huan-Qiang Zhou}
\affiliation{Centre for Modern Physics and Department of Physics,
Chongqing University, Chongqing 400044, The People's Republic of
China}

\begin{abstract}
The Kosterlitz-Thouless transition is studied from the
representation of the systems's ground state wave functions in terms
of Matrix Product States for a quantum system on an infinite-size
lattice in one spatial dimension.   It is found that, in the
critical regime for a one-dimensional quantum lattice system with
continuous symmetry, the newly-developed infinite Matrix Product
State algorithm automatically leads to infinite degenerate ground
states, due to the finiteness of the truncation dimension. This
results in \textit{pseudo} continuous symmetry spontaneous
breakdown, which allows to introduce a pseudo-order parameter that
must be scaled down to zero, in order to be consistent with the
Mermin-Wegner theorem. We also show that the ground state fidelity
per lattice site exhibits a \textit{catastrophe point}, thus
resolving a controversy regarding whether or not the ground state
fidelity is able to detect the Kosterlitz-Thouless transition.

\end{abstract}


\maketitle

 {\it Introduction.} The Kosterlitz-Thouless (KT) transition~\cite{kt}, an infinite order
transition from a critical phase to a gapful phase, is ubiquitous in
quantum systems on an infinite-size lattice in one spatial
dimension. It describes one of the instabilities of the Luttinger
liquid induced by an marginal perturbation. Remarkably, it does not
arise from the long-range ordering, thus it falls outside of the
conventional Landau-Ginzburg-Wilson paradigm, in which the most
fundamental notion is spontaneous symmetry breaking
(SSB)~\cite{anderson,coleman}, with the symmetry-broken phase
characterized by a nonzero local order parameter. Indeed, the type
of internal order that a system possesses is profoundly affected by
its dimensionality. In quantum field theory, the Mermin-Wegner
theorem states that continuous symmetries cannot be spontaneously
broken in systems with local interactions in one spatial
dimension~\cite{mw}. That is, in quantum systems with a continuous
symmetry in one spatial dimension, a genuine long-range order is
destroyed by quantum fluctuations. Instead, a quasi-long-range order
occurs in the critical phase, which is characterized by a power-law
decay in correlation functions.

The peculiarity of the KT transition makes it very difficult to map
out the ground state phase diagrams for various quantum lattice
systems in one spatial dimension. Indeed, a numerical analysis of
the KT transition suffers from pathological problems. One of these
problems is that the finite size scaling technique~\cite{fss}, which
is successful for second order quantum phase transitions
(QPTs)~\cite{sachdev,wen}, can not be applied to the KT
transition~\cite{np}, since there are logarithmic corrections from
the marginal perturbation. This motivated Nomura and
Okamoto~\cite{no} to develop the so-called level spectroscopy to
overcome these difficulties by combining the renormalization group,
conformal field theory and the symmetry. This raises an intriguing
question as to whether or not there is a unifying practical approach
to different types of QPTs.

In this Letter, we address this issue from a novel perspective. The
new inputs come from recent advances in both classical simulations
of quantum lattice systems and our understanding of QPTs. First, a
tensor network (TN) representation of quantum many-body wave
functions provides an efficient way to classically simulate quantum
many-body systems, which include Matrix Product States (MPS) in one
spatial dimension and the Projected Entangled-Pair States (PEPS) in
two or more spatial dimensions. Second, two novel approaches to
study QPTs have been proposed in terms of both
entanglement~\cite{amico} and
fidelity~\cite{zp,zhou,zov,fidelity,mfyang,fidelity1}. In
Refs.~\cite{zhou,zov}, it was argued that the ground state fidelity
per lattice site, combining with a practical way to compute it using
the TN algorithms for translationally invariant infinite-size
systems, is able to capture the  many-body physics underlying
various quantum lattice systems in condensed matter.

Although an MPS is well suited for the description of a gapped
quantum state, it remains mysterious how it works for quantum states
in the gapless regime. This casts doubts on the applicability of the
NT algorithms to the study of the KT transition. Our approach is
based on the observation that, in the gapless regime, the
newly-developed infinite MPS (iMPS) algorithm automatically leads to
infinite degenerate ground states, due to the finiteness of the
truncation dimension. This results in \textit{pseudo} continuous
SSB~\cite{pseudossb}, which allows to introduce a pseudo-order
parameter that must be scaled down to zero, in order to be
consistent with the Mermin-Wegner theorem. We also show that the
ground state fidelity per lattice site exhibits a
\textit{catastrophe point}~\cite{bifurcation} at a pseudo critical
point for any finite truncation dimension. Normally, an
extrapolation to infinite truncation dimension should be performed
to determine the KT transition point. This resolves a controversy
regarding whether or not the ground state fidelity is able to detect
the KT transition~\cite{mfyang,fidelity1}.

{\it Matrix Product States on an infinite-size lattice in one
spatial dimension and continuous symmetries.} The iMPS algorithm is
a variational algorithm to compute the MPS representations of ground
state wave functions for translationally invariant quantum systems
on an infinite-size lattice in one spatial dimension~\cite{vidal1}.
Assume that the Hamiltonian takes the form: $H =\sum _i h^{[i,
i+1]}$, with $h^{[i, i+1]}$ being the nearest-neighbor two-body
Hamiltonian density.  The problem for finding the system's ground
state wave functions amounts to computing the imaginary time
evolution for a given initial state $|\Psi (0)\rangle |$: $ |\Psi
(\tau)\rangle = \exp(- H \tau) |\Psi (0)\rangle / |\exp(- H \tau)
|\Psi (0)\rangle |$. An efficient way to do this is based on the
observation that the imaginary time evolution operator is reduced to
a product of two-site evolution operators acting on sites $i$ and
$i+1$: $U(i,i+1) = \exp(-h^{[i,i+1)]} \delta \tau )$, $\delta \tau
<<1$, as follows from the Suzuki-Trotter
decomposition~\cite{suzuki}, and the fact that any wave function
admits an MPS representation in a canonical form: attached to each
site is a three-indices tensor $\Gamma^s_{A\;lr}$ or
$\Gamma^s_{B\;lr}$, and to each bond a diagonal matrix $\lambda_A$
or $\lambda_B$, depending on the evenness and oddness of the $i$-th
site and the $i$-th bond, respectively. Here, $s$ is a physical
index, $s=1,\cdots,d$, with $d$ being the dimension of the local
Hilbert space, and $l$ and $r$ denote the bond indices,
$l,r=1,\cdots, \chi$, with $\chi$ being the truncation dimension.
The effect of a two-site gate $U(i,i+1)$ may be absorbed by
performing a singular value decomposition. This allows to update the
tensors involved in the MPS representation. Repeating this procedure
until the ground state energy converges, one may generate the
system's ground state wave functions in the MPS representation.

It is known that the iMPS algorithm yields the best approximation to
the ground state wave function (for a fixed truncation dimension
$\chi$), as long as the ground state is gapful.  However, it is
surprising to see that it also works for quantum lattice systems
with continuous symmetries in a gapless regime.  The key point here
is that continuous symmetries and the translational invariance under
one site shifts cannot be maintained simultaneously during the
imaginary time evolution~\cite{af}. Indeed, in order to mimic the
gaplessness of excitations in the gapless regime, the iMPS algorithm
resorts to the Goldstone mode. That is, the best approximation to
the ground state wave function is \textit{not unique}, if the
translational invariance under one site shifts is
maintained~\cite{sukhi}. Instead, infinite degenerate ground states
are generated, each of which breaks the continuous symmetry. From
now on, we restrict ourselves to the symmetry group $U(1)$.

{\it Pseudo continuous symmetry spontaneous breakdown and
pseudo-order parameters.} For a $U(1)$ invariant quantum system on
an infinite-size lattice in one spatial dimension, the iMPS
algorithm automatically produces infinite degenerate ground states,
each of which breaks the $U(1)$ symmetry. Moreover, the symmetry
breakdown results from the fact that an initial state has been
chosen randomly. That is, a phenomenon occurs which shares all the
features of an SSB~\cite{anderson,coleman}. Indeed, the implication
of an SSB is two-fold: first, a system has stable and degenerate
ground states, each of which breaks the symmetry of the system;
second, the symmetry breakdown results from random perturbations. In
addition, such a ``symmetry-breaking order" may be quantified by
introducing a local ``order parameter", which may be read off from
the reduced density matrix on a local area~\cite{zhou3}. However,
this is in apparent contradiction with the Mermin-Wegner theorem,
which states that continuous symmetries cannot be spontaneously
broken for quantum systems in one spatial dimension~\cite{mw}.  To
resolve this contradiction, one has to require that the ``order
parameter'' must be scaled down to zero, when the truncation
dimension $\chi$ tends to $\infty$. In order to distinguish them
from a genuine SSB and a local order parameter, we introduce the
notions of a \textit{pseudo} SSB and a pseudo-order parameter. In
this scenario, the Goldstone mode survives as gapless excitations in
the critical phase, and the KT transition is a limiting case of the
conventional symmetry-breaking order, when the truncation dimension
goes to $\infty$.

{\it Ground state fidelity per lattice site.} Suppose that a $U(1)$
invariant quantum lattice system undergoes the KT transition, when a
control parameter $\lambda$ crosses a critical point $\lambda_c$.
Let us see whether or not the ground state fidelity per lattice site
is able to detect it.  The ground state fidelity per lattice site,
$d(\lambda_1, \lambda_2)$, is defined as the scaling parameter,
which characterizes how fast the fidelity $F(\lambda_1, \lambda_2)
\equiv |\langle \Psi (\lambda_2) |\Psi (\lambda_1) \rangle | $
between two ground states $|\Psi (\lambda_1) \rangle$ and $|\Psi
(\lambda_2) \rangle$ goes to zero when the thermodynamic limit is
approached~\cite{zhou,zov}. In fact, the ground state fidelity
$F(\lambda_1, \lambda_2)$ asymptotically scales as $F(\lambda_1,
\lambda_2) \sim d(\lambda_1, \lambda_2)^L$, with $L$ the number of
sites in a finite-size lattice. Remarkably, $d(\lambda_1,
\lambda_2)$ is well defined in the thermodynamic limit, and
satisfies the properties inherited from the fidelity $F(\lambda_1,
\lambda_2)$: (i) normalization $d(\lambda, \lambda) =1$; (ii)
symmetry $d(\lambda_1, \lambda_2) = d(\lambda_2, \lambda_1)$; and
(iii) range $0 \le d(\lambda_1, \lambda_2) \le 1$.

In the $U(1)$ symmetric phase, the ground state is
non-degenerate~\cite{kttype}, whereas in the $U(1)$ symmetry-broken
phase, infinite degenerate ground states arise. This implies that,
if we choose $\Psi (\lambda_2)$ as a reference state, with
$\lambda_2$ in the $U(1)$ symmetric phase, then  the ground state
fidelity per lattice site, $d(\lambda_1, \lambda_2)$, cannot
distinguish infinite degenerate ground states in the $U(1)$
symmetry-broken phase. This follows from the fact that $\langle \Psi
(\lambda_2) |\Psi_\eta (\lambda_1) \rangle = \langle \Psi
(\lambda_2) |U(\xi) |\Psi_\eta (\lambda_1) \rangle = \langle \Psi
(\lambda_2) |\Psi_{\eta+\xi} (\lambda_1) \rangle$, for any large but
finite size $L$, with $U(\xi)$ being an element of the symmetry
group $U(1)$, and the subscript $\eta$ in $ |\Psi_\eta (\lambda)
\rangle$ labeling the eigenstates of the $U(1)$ generator. However,
if we choose $\Psi (\lambda_2)$ as a reference state, with
$\lambda_2$ in the $U(1)$ symmetry-broken phase, then $d(\lambda_1,
\lambda_2)$ is able to distinguish infinite degenerate ground
states. Therefore, the phase transition point $\lambda_c$ manifests
itself as a \textit{catastrophe  point} for any \textit{finite}
$\chi$. The catastrophe  point disappears as $\chi$ tends to
$\infty$. However, the critical point $\lambda_c$ may be determined
by performing a scaling analysis with reasonably small
$\chi$'s~\cite{vneumann}.

{\it Examples.} The first example we shall investigate is the spin
1/2 XXZ model described by the Hamiltonian:
\begin{equation}
\small
  H=\sum_{i=-\infty}^\infty \left(S^{[i]}_{x}S^{[i+1]}_{x}+
  S^{[i]}_{y}S^{[i+1]}_{y}+ \Delta S^{[i]}_{z}S^{[i+1]}_{z}\right), \label{ham2}
\end{equation}
where $S^{[i]}_{\alpha}$ ($\alpha=x,y,z$) are the Pauli spin
operators at the site $i$, and $\Delta$ denotes the anisotropy in
the internal spin space. The model is in the critical regime (CR)
for $\Delta \in (-1, 1]$, with the same universality class as a free
bosonic field theory, and exhibits anti-ferromagnetic (AF) and
ferromagnetic (FM) orders, respectively, for $\Delta > 1$ and
$\Delta < -1$. Besides a continuous $U(1)$ symmetry, the model also
enjoys a $Z_2$ symmetry,  generated by the operation:
$S^{[i]}_{x}\leftrightarrow S^{[i]}_{y}$ and $S^{[i]}_{z}\rightarrow
-S^{[i]}_{z}$.

The second example is the spin 1 XXZD model with uniaxial single-ion
anisotropy:
\begin{equation}
\small
 H=\sum_{i=-\infty}^\infty \left(S^{[i]}_{x}S^{[i+1]}_{x}+
  S^{[i]}_{y}S^{[i+1]}_{y}+J_z S^{[i]}_{z}S^{[i+1]}_{z}\right)+ D \sum_{i=-\infty}^\infty  (S^{[i]}_{z})^2, \label{ham3}
\end{equation}
where $S^{[i]}_{\alpha}$ ($\alpha=x,y,z$) are the spin 1 operators
at the lattice site $i$, and $J_z$ and $D$ are the Ising-like and
single-ion anisotropies, respectively.  For $J_z =-1/2$, there are
two phase transitions, i.e., the KT transition from the large-$D$
(LD) to a CR at $D \sim0.693 $, and a first order QPT from the CR to
a FM phase at $D \sim-1.184 $, as $D$ varies from $\infty$ to
$-\infty$~\cite{xxzd}.

\begin{table}
\centering
\begin{tabular}{|c|p{0.25cm}|p{1.25cm}|p{1.25cm}|p{1.25cm}|p{1.25cm}|p{1.25cm}|}
\hline
Model   &\hfil $\alpha$ \hfil & \hfil $\beta$\hfil & \hfil $\gamma$\hfil & \hfil $\delta$\hfil & \hfil $\mu$\hfil & \hfil $\nu$\hfil  \\[1pt] \cline{1-7}
XXZ & \hfil  0 \hfil&\hfil -0.00186 \hfil & \hfil 0.17359 \hfil &
\hfil $-$\hfil & \hfil $-$\hfil & \hfil $-$\hfil  \\ [1pt]
\cline{2-7}
 spin 1/2 & \hfil   0 \hfil&\hfil   0.16882 \hfil & \hfil 0.04045 \hfil & \hfil $-$\hfil & \hfil $-$\hfil & \hfil $-$\hfil  \\ [1pt] \cline{1-7}
XXZD &\hfil  0 \hfil & \hfil  -0.24659 \hfil & \hfil -0.27287 \hfil
& \hfil 0.00314 \hfil & \hfil 0.08843 \hfil & \hfil  -0.00204 \hfil
\\ [0pt] \cline{2-7}
 spin 1 & \hfil   0 \hfil&\hfil -0.24659 \hfil & \hfil -0.08133 \hfil & \hfil 0.26049 \hfil & \hfil -0.07274 \hfil & \hfil -0.05033 \hfil  \\ [0pt]\cline{1-7}
\hline
\end{tabular}

\setlength{\abovecaptionskip}{5pt} \caption{The one-site reduced
density matrices for the spin $1/2$ XXZ model ($\Delta =0$) and the
spin 1 XXZD model with uniaxial single-ion anisotropy ( $J_z=-0.5,
D=-0.3$): $ \rho^{[i]}_{XXZ}= \frac{1}{2} I +\alpha S^{[i]}_{z}+
\beta S^{[i]}_{x}+ \gamma  S^{[i]}_{y}$ and
$\rho^{[i]}_{XXZD}=\frac{1}{3}(1-2\beta) I +\alpha S^{[i]}_{z}+
\beta ( S^{[i]}_{z} )^{2}+\sqrt{2}\gamma S^{[i]}_{x}+\sqrt{2} \delta
S^{[i]}_{y}+ \mu ((S^{[i]}_{x})^{2}-(S^{[i]}_{y})^{2})+
\nu(S^{[i]}_{x} S^{[i]}_{y} +S^{[i]}_{y} S^{[i]}_{x})$. The two
degenerate ground states originated from two randomly chosen initial
states are connected via $S^{[i]'}_x =\cos \theta S^{[i]}_x + \sin
\theta S^{[i]}_y, S^{[i]'}_y =- \sin \theta S^{[i]}_x + \cos \theta
S^{[i]}_y$, with $\theta =77.140 $ for the spin $1/2$ XXZ model and
$\theta=72.001$ for the spin 1 XXZD model. } \label{TAB1}
\end{table}

{\it Simulation results.} In Table~\ref{TAB1}, we present the
specific coefficients in the one-site reduced density matrices for
the spin $1/2$ XXZ model ($\Delta =0$) and the spin 1 XXZD model
with uniaxial single-ion anisotropy ( $J_z=-0.5, D=-0.3$): $
\rho^{[i]}_{XXZ}= \frac{1}{2} I +\alpha S^{[i]}_{z}+ \beta
S^{[i]}_{x}+ \gamma S^{[i]}_{y}$ and
$\rho^{[i]}_{XXZD}=\frac{1}{3}(1-2\beta) I +\alpha S^{[i]}_{z}+
\beta (S^{[i]}_{z})^{2}+\sqrt{2}\gamma S^{[i]}_{x}+\sqrt{2} \delta
S^{[i]}_{y}+ \mu ((S^{[i]}_{x})^{2}-(S^{[i]}_{y})^{2})+
\nu(S^{[i]}_{x} S^{[i]}_{y} +S^{[i]}_{y} S^{[i]}_{x})$. The two
degenerate ground states originated from two randomly chosen initial
states are connected via $S^{[i]'}_x =\cos \theta S^{[i]}_x + \sin
\theta S^{[i]}_y, S^{[i]'}_y =- \sin \theta S^{[i]}_x + \cos \theta
S^{[i]}_y$, with $\theta =77.140 $ for the spin $1/2$ XXZ model (the
fidelity per lattice site between them is $0.9976$) and
$\theta=72.001$ for the spin 1 XXZD model (the fidelity per lattice
site between them is $0.9925$). The iMPS simulations are performed
for a randomly chosen initial state, with $\chi$ to be 16.

\begin{figure}
 \includegraphics[width=0.72\textwidth]{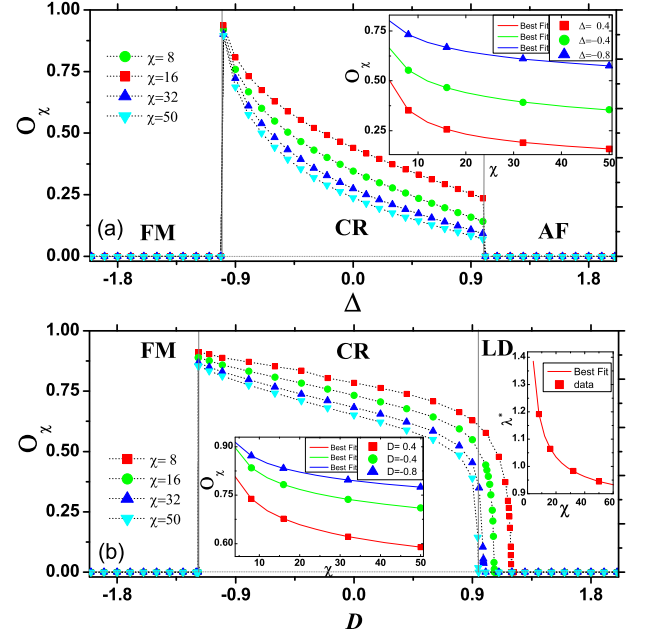}

\caption{(color online) Main: (a) The magnitude $O_\chi$ of the
pseudo-order parameter $(\langle S^{[i]}_x \rangle, \langle
S^{[i]}_y \rangle)$ for the spin $1/2$ XXZ model and (b) that for
the spin 1 XXZD model (with a fixed $J_z =-1/2$). The iMPS
simulations are performed for a randomly chosen initial state, with
the truncation dimension $\chi$ to be 8, 16, 32, and 50. Note that
the pseudo-order parameter is discontinuous even at $\Delta =1$ for
the XXZ model. However, this is an artifact of the iMPS algorithm.
Inset: The pseudo-order parameter is scaled down to zero, according
to $O_\chi = a \chi^{-b}(1+c \chi^{-1})$, in the critical phase, in
the left-top inset in (a) for the XXZ model (with $a =0.7331, 0.8766
,0.9545, b=0.3947, 0.2327, 0.1296, c=0.7493, 0.1898, 0.0224$ for
$\Delta= 0.4, -0.4, -0.8$, respectively), and the middle-bottom
inset in (b) for the XXZD model (with $a=0.9454, 0.9725, 0.9887,
b=0.1212, 0.0811, 0.0625, c=0.0405, 0.1314, 0.0322$ for $D= 0.4,
-0.4, -0.8$, respectively). Moreover, an extrapolation to $\chi
=\infty$ is performed for the pseudo transition points (at which the
pseudo-order parameter becomes zero), yielding the KT transition
point ($D=0.827$) for the XXZD model in the left-top inset in (b).}
\label{FIG1}
\end{figure}

We may choose $(\langle S^{[i]}_x \rangle, \langle S^{[i]}_y
\rangle)$ as the pseudo-order parameter for both the
models~\cite{qop}. In Fig.~\ref{FIG1}\;(a) and (b), we plot the
magnitude $O_\chi = \sqrt {\langle S^{[i]}_x \rangle ^2 + \langle
S^{[i]}_y \rangle ^2}$ of the pseudo-order parameter for the spin
$1/2$ XXZ model and the spin 1 XXZD model (with a fixed $J_z
=-1/2$), respectively.  The iMPS simulations are performed for a
randomly chosen initial state, with the truncation dimension $\chi$
to be 8, 16, 32, and 50. This indicates that  a first order QPT and
the KT transition occur at $\Delta =-1$ and $\Delta =1$,
respectively, for the spin $1/2$ XXZ model, and that a first order
QPT and the KT transition occur at $D =-1.184$ and $D =0.827$,
respectively, for the spin 1 XXZD model. Note that an extrapolation
to $\chi =\infty$ has been performed for the pseudo transition
points (at which the pseudo-order parameter becomes zero), yielding
the KT transition point ($D=0.827$) for the XXZD model (see the
left-top inset in Fig.~\ref{FIG1}\;(b)), which is larger compared to
$D \sim0.693 $ from the level spectroscopy~\cite{xxzd}. The
pseudo-order parameter is scaled down to zero in the critical phase,
as shown in the left-top inset in Fig.~\ref{FIG1}\;(a) for the XXZ
model, and the middle-bottom inset in Fig.~\ref{FIG1}\;(b) for the
XXZD model, to keep consistent with the Mermin-Wegner theorem.

\begin{figure}
\includegraphics[width=0.72\textwidth]{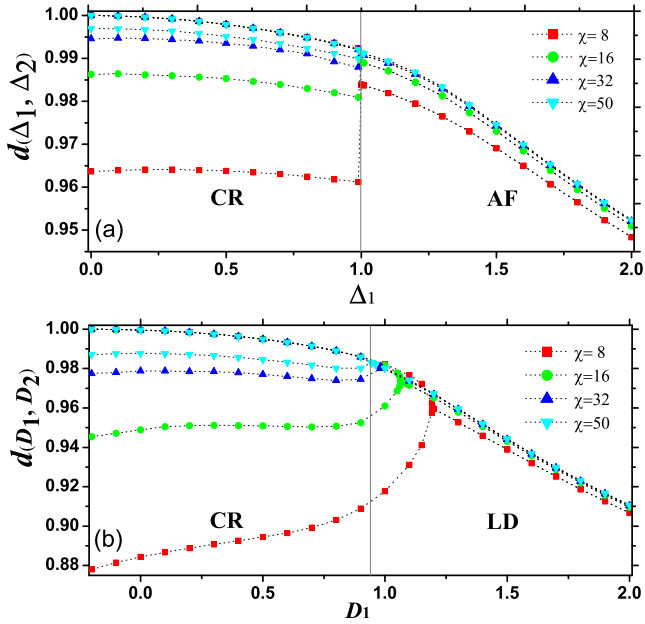}
 \caption{(color online)  The ground state fidelity per lattice
site, $d(\Delta_1, \Delta_2)$, for  the spin $1/2$ XXZ model (a)
and, $d(D_1, D_2)$, for the spin 1 XXZD model (with a fixed $J_z
=-1/2$) (b). Here, the truncation dimension $\chi$ takes 8, 16, 32,
and 50, respectively. We have chosen $\Psi (\Delta_2)$ ($\Psi
(D_2)$) as a reference state, with $\Delta_2$ ($D_2$) in the $U(1)$
symmetry-broken phase $\Delta_2 =0$ ($D_2 =-0.2$), then $d(\Delta_1,
\Delta_2)$ ($d(D_1, D_2)$ ) is able to distinguish infinite
degenerate ground states, with a (pseudo) phase transition point as
a \textit{catastrophe point}. Note that $d(\Delta_1, \Delta_2)$ is
discontinuous even at $\Delta_2 =1$ for the XXZ
model~\cite{comment}.}
  \label{FIG2}
   \end{figure}

In Fig.~\ref{FIG2}\;(a) and (b), we plot the ground state fidelity
per lattice site, $d(\Delta_1, \Delta_2)$ , for  the spin $1/2$ XXZ
model~\cite{luttinger} and the spin 1 XXZD model (with a fixed $J_z
=-1/2$). We have chosen $\Psi (\Delta_2)$ ($\Psi (D_2)$) as a
reference state, with $\Delta_2$ ($D_2$) in the $U(1)$
symmetry-broken phase $\Delta_2 =0$ ($D_2 =-0.2$), then $d(\Delta_1,
\Delta_2)$ ($d(D_1, D_2)$ ) is able to distinguish infinite
degenerate ground states, with a (pseudo) phase transition point as
a \textit{catastrophe point} (which coincides with the pseudo
transition point from the pseudo-order parameter for a given $\chi$,
within the accuracy). Note that $d(\Delta_1, \Delta_2)$ is
discontinuous even at $\Delta_2 =1$ for the XXZ
model~\cite{comment}. However, this is just an artifact of the iMPS
algorithm, since all the degenerate ground states collapse into the
genuine ground state as $\chi \rightarrow \infty$.

We have also computed  the von Neumann entropy for the spin $1/2$
XXZ model and the spin 1 XXZD model (with a fixed $J_z =-1/2$) (not
shown here).  It detects the QPTs~\cite{tag}, but does not
distinguish infinite degenerate ground states arising from a pseudo
SSB for a given $\chi$.


{\it Acknowledgements.} The work is supported by the National
Natural Science Foundation of China (Grant Nos: 10774197 and
10874252) and the Natural Science Foundation of Chongqing (Grant No:
CSTC, 2008BC2023).



\begin{thebibliography}{10}
\bibitem{pseudossb} The term ``pseudo SSB" describes a
numerical phenomena, in which an SSB is induced by the truncation in
the iMPS algorithm. It vanishes when the truncation is removed.
\bibitem{af} For an antiferromagnetic spin system, the translational invariance under
one site shifts should be accompanied by the spin reversal at all
the sites simultaneously.
\bibitem{vneumann} It is possible to locate the critical point from
the von Neumann entropy, a bipartite entanglement measure. However,
the von Neumann entropy is fully determined by the singular value
matrices $\lambda_A$ and $\lambda_B$, which are the same due to the
fact that the translational invariance under one site shifts is
maintained. Because all the information concerning a pseudo SSB is
encoded in $\Gamma^s_{A\;lr}$ and $\Gamma^s_{B\;lr}$, the von
Neumann entropy fails to detect a pseudo SSB.
\bibitem{qop} Another choice of the pseudo-order parameter for the
spin 1 XXZD model is $(\langle (S^{[i]}_x)^2 - (S^{[i]}_y)^2
\rangle, \langle S^{[i]}_x S^{[i]}_y+ S^{[i]}_y S^{[i]}_x \rangle)$.
\bibitem{luttinger} Our simulation result of $d(\Delta_1, \Delta_2)$ for the XXZ model does not
match the exact expression $\ln d(\Delta_1, \Delta_2)=1/2 (\ln 2-
\ln (\sqrt {K(\Delta_1)/K(\Delta_2)} + \sqrt
{K(\Delta_2)/K(\Delta_1))}$, derived from the Luttinger liquid
description, combining with the Bethe ansatz: $K(\Delta)=\pi /(2(\pi
- \arccos \Delta))$~\cite{mfyang}. This is expected, since the
Luttinger liquid only describes the low energy physics for the XXZ
model, whereas $d(\Delta_1, \Delta_2)$ captures the physics in both
the low energy and high energy sectors.
\bibitem{comment} For first order QPTs, the ground state
fidelity per lattice site is discontinuous~\cite{zhou,zov}.
\end{thebibliography}
\end{document}